\documentclass[twocolumn,nofootinbib]{nature}

\usepackage{amsmath}    
\usepackage{color}
\usepackage[10pt]{moresize}

\usepackage{graphicx}
\usepackage{caption}
\DeclareCaptionLabelFormat{adja-page}{\hrulefill\\#1 #2 \emph{(previous page)}}
\usepackage{subfig}

\usepackage{amsfonts}
\usepackage{amssymb}
\usepackage{amscd}
\usepackage{enumerate}
\usepackage{epsfig}
\usepackage{epstopdf}
\usepackage{dcolumn}
\usepackage{bm}
\usepackage{amsthm}
\usepackage{amsfonts}
\usepackage{color} 
\usepackage[usenames,dvipsnames]{xcolor}
\usepackage{amsmath}
\usepackage{enumerate}
\usepackage{bbm}
\usepackage[colorlinks=true,citecolor=blue,urlcolor=black]{hyperref}

\begin{document}

\title{Experimental round-robin differential phase-shift quantum key distribution}

\author{Yu-Huai Li$^{1,2,\ast}$, Yuan Cao$^{1,2,\ast}$, Hui Dai$^{1,2}$, Jin Lin$^{1,2}$, Zhen Zhang$^{3}$, Wei Chen$^{1,2}$, Yu Xu$^{1,2}$, Jian-Yu Guan$^{1,2}$, Sheng-Kai Liao$^{1,2}$, Juan Yin$^{1,2}$, Qiang Zhang$^{1,2}$, Xiongfeng Ma$^{3}$, Cheng-Zhi Peng$^{1,2}$, Jian-Wei Pan$^{1,2}$}

\maketitle

\begin{affiliations}
\item Shanghai Branch, National Laboratory for Physical Sciences at Microscale and Department of Modern Physics, University of Science and Technology of China, Shanghai 201315, China.
\item Synergetic Innovation Center of Quantum Information and Quantum Physics, University of Science and Technology of China, Hefei, Anhui 230026, China.
\item Center for Quantum Information, Institute for Interdisciplinary Information Sciences, Tsinghua University, Beijing 100084, China.

$^\ast$ These authors contributed equally to the paper
\end{affiliations}

\baselineskip24pt

\maketitle

\begin{abstract}
In conventional quantum key distribution (QKD) protocols, security is guaranteed by estimating the amount of leaked information through monitoring signal disturbance, which, in practice, is generally caused by environmental noise and device imperfections rather than eavesdropping. Such estimation therefore tends to overrate the amount of leaked information in practice, leads to a fundamental threshold of the bit error rate. The threshold becomes a bottleneck of the development of practical QKD systems. In classical communication, according to Shannon's communication theory, information can transform through a noisy channel even if the background noise is very strong compare to the signal and hence the threshold of the bit error rate tends to $50\%$\cite{shannon1948mathematical}. One might wonder whether a QKD scheme can also tolerate error rate as high as $50\%$. The question is answered affirmatively with the recent work of round-robin differential phase-shift (RRDPS) protocol\cite{sasaki2014practical,zhang2015round}, which breaks through the fundamental threshold of the bit error rate and indicates another potential direction in the field of quantum cryptography. The key challenge to realize the RRDPS scheme lies on the measurement device, which requires a variable-delay interferometer. The delay needs to be chosen from a set of predetermined values randomly. Such measurement can be realized by switching between many interferometers with different delays at a high speed in accordance with the system repetition rate. The more delay values can be chosen from, the higher error rate can be tolerated. By designing an optical system with multiple switches and employing an active phase stabilization technology, we successfully construct a variable-delay interferometer with 128 actively selectable delays. With this measurement, we experimentally demonstrate the RRDPS QKD protocol and obtain a final key rate of $15.54$ bps via a total loss of 18 dB and 8.9\% error rate. Our experiment demonstrates that the RRDPS QKD protocol is practical with current technology.
\end{abstract}


In a QKD protocol, the sender, Alice, sends a quantum signal through an untrusted quantum channel. The receiver, Bob, measures the arrived quantum signal and obtains the raw key. Because of the device imperfections, environmental interference and possibly eavesdropping, the raw keys of Alice and Bob maybe not be identical or private. In the security analysis, the disturbance of the signal is quantified by the bit flip error rate, $e_{bit}$, and the leaked information is quantified by the phase error rate, $e_{ph}$. To ensure the final keys are identical and secure, a proper postprocessing \cite{Lo1999Science,SHORPRESKILL2000PRL} should be performed. After performing the error correction, which removes the disturbances, and one should apply privacy amplification, which removes the leaked information. The final key generation rate per raw key bit is given by\cite{SHORPRESKILL2000PRL},
\begin{equation}\label{Generalkeyrate}
R = 1-H(e_{bit}) - H(e_{ph}).
\end{equation}
When the error rates, $e_{bit}$ and $e_{ph}$, exceeds some thresholds, the key rate becomes zero or negative and hence no secure keys can be generated.

In conventional QKD protocols, security is guaranteed by the Heisenberg uncertainty principle which guarantees that any attempts of eavesdropping the quantum channel would inevitably cause quantum signal disturbances. Therefore, the leaked information obtained by an eavesdropper, Eve, can be upper-bounded by the disturbance of the signal. In the Bennett-Brassard-1984 (BB84) protocol, due to its symmetry, the phase error rate is estimated by the bit error rate $e_{ph}= e_{bit}$ and hence the final key rate, in Eq.~\eqref{Generalkeyrate}, is given by $R = 1-2H(e_{bit})$. When $e_{bit}>11\%$, the final key rate approaches to $0$. Hence, the bit error rate threshold for the BB84 protocol is $11\%$ using the above postprocessing. Note that with other postprocessing techniques\cite{1176619}, a higher bit error rate threshold can be obtained. Nevertheless, there exists a fundamental limitation on the error rate threshold\cite{PhysRevLett.92.217903}. It is widely believed that secure QKD cannot be achieved when the background noise is too strong compare to the signal. In practice, the strength of background noise can be considered as a constant, whereas the strength of the signal exponentially decreases with an increasing transmission distance. Hence, the error rate threshold puts a fundamental limit on the secure transmission distance. From a realistic point of view, most of disturbances are caused by the environment noises and device imperfections instead of eavesdropping. Thus the leaked information is often overestimated, which is the root for the limit on the error rate threshold.

In classical communication, according to Shannon's communication theory, such error rate limit does not exist as long as $e_{bit}\neq50\%$\cite{shannon1948mathematical}. One might wonder whether a QKD scheme can also tolerate error rate as high as $50\%$. Recently, a new QKD protocol\cite{sasaki2014practical} called RRDPS was proposed, which is essentially evolved from the differential phase-shift (DPS) protocol\cite{IWY_DPS03,Takesue:40dBQKD:2007}. Surprisingly, with the new protocol, secure key can be generated even if the bit flip error rate close to $50\%$.

\begin{figure}[!t]\center
\resizebox{16cm}{!}{\includegraphics{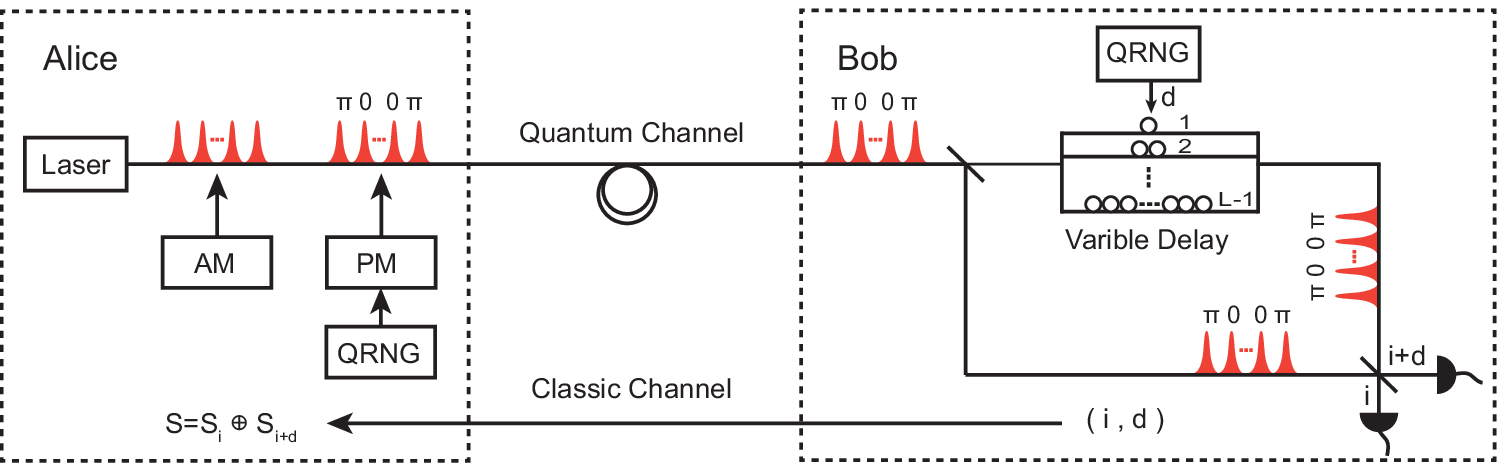}}
\caption{\textbf{Schematic diagram of RRDPS QKD proposal.} \textbf{State preparation:} Alice generates an $L$-pulse train by modulating a narrow linewidth continuous-wave (CW) laser with an amplitude modulator (AM). Then random phase shift of $0$ or $\pi$ is performed in each pulse. After attenuating the intensity of an $L$-pulse train to a preset $\nu$, the L-pulse train is transmitted to Bob through a fibre channel.
\textbf{Measurement:} Bob generates two random numbers: $c$ from $\{0,1\}$, which denotes the direction of the delay, and $d$ from $\{1,2,\dots,L-1\}$, which denotes the length of the delay. Then the arrived pulse train is split into two with a beamsplitter (BS) and a random delay of $r^\prime=(-1)^cd$ pulses is applied to one of the split trains.  Then the two pulse trains are combined on a beam splitter, performed an interference measurement and obtained a index, $i$,of the pulse which is detected a photon together with the sifted key $s_{i}\oplus s_{(i+r^\prime\mod L)}$. \textbf{Announcement:} Bob announces $i$ and $(i+r\mod L)$ so that Alice can calculate the sifted key $s_{i}\oplus s_{(i+r^\prime \mod L)}$.}
\label{Fig:Schematic}
\end{figure}

The schematic diagram of the RRDPS QKD scheme is shown in Fig.~\ref{Fig:Schematic}. Alice prepares a pulse train containing $L$ pulses, encodes the (random) key information into the phase of each pulse, $0$ or $\pi$, and sends it to Bob, who splits the $L$-pulse train into two with a beam splitter. Bob randomly shifts (backward or forward) one of the split pulse trains by $d$ pulses where $1\le d\le L-1$, and interferes them. The detection result that Bob obtains essentially reveals the phase difference between two pulses $i$ and $i\pm d$, where $+$ stands for shifting forward and $-$ stands for backward. The key rate of the RRDPS QKD protocol is given by\cite{zhang2015round}
\begin{equation}\label{eq:extGLLP}
\begin{aligned}
R &= Q\left[1-H(e_{\textrm{bit}})-H_{\mathrm{PA}}\right],
\end{aligned}
\end{equation}
where $ R$ is the final key bit per $L$-pulse train. In experiment, the average number of valid detections per $L$-pulse train, $Q$, can be measured directly. In Method, we show how to estimate the privacy amplification term $H_{\mathrm{PA}}$.

\begin{figure}[!t]\center
\resizebox{16cm}{!}{\includegraphics{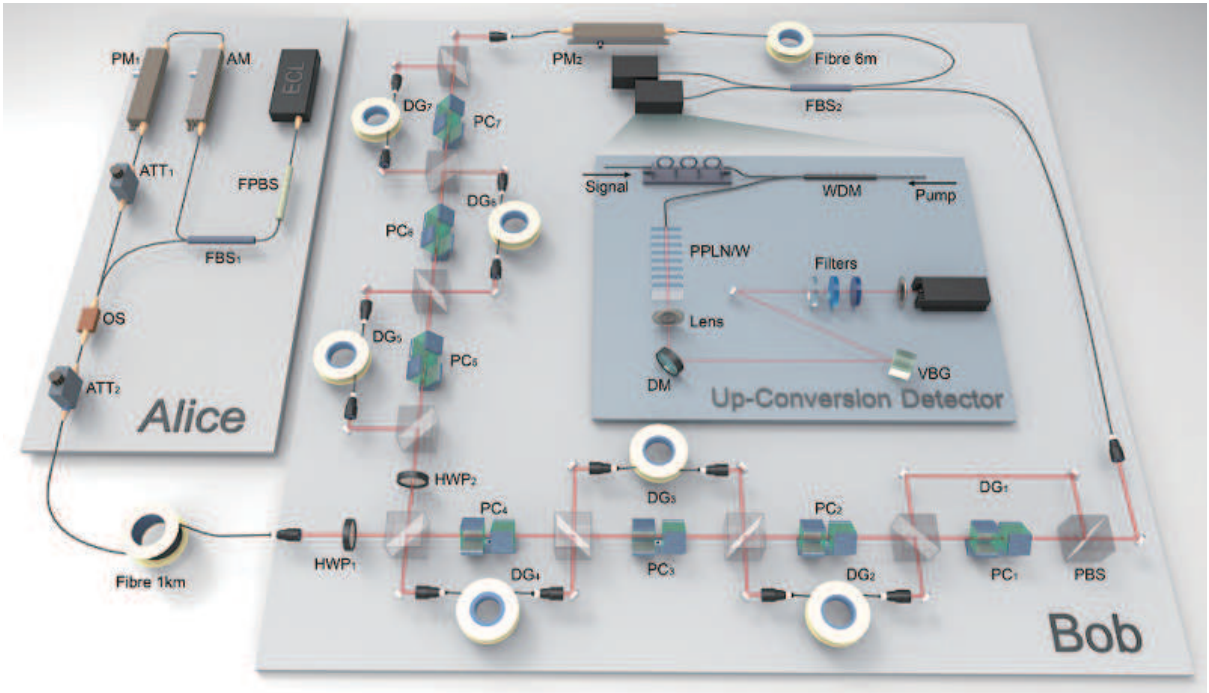}}
\caption{\textbf{Experimental setup.} ECL: external cavity laser; AM: amplitude modulator; PM: phase modulator; ATT: attenuator; (F) PBS: (fibre) polarizing beam splitter; FBS: fibre beam splitter; OS: optical switch; HWP: half-wave plate; PC: pockels cell; DG: delay gate; WDM: wavelength division multiplexing. The 7 delay gates are implemented by 6 fibre links with appropriate length and a free-space link. 7 pockels cells are deployed to control the state of each delay gate. The interference measurement is performed on $FBS_2$ and completed on two up-conversion detectors, whose details are shown in the middle of Bob's setup.}
\label{Fig:Setup}
\end{figure}

In this letter, we demonstrate the RRPDS QKD protocol with $L=128$. The setup is shown in Fig.~\ref{Fig:Setup}. On sender side, a continuous-wave (CW) external cavity laser (ECL) is employed as the optical source. The central wavelength of the ECL is $1550.12nm$ with a linewidth below $2kHz$, which can provide a coherence time beyond $500\mu s$. This CW laser is modulated by an amplitude modulator (AM, Photline 10GHz) in order to produce a $128$-pulse train. The pulses, with a full width at half maximum (FWHM) of $300ps$, are separated by $2ns$. Thus the overall duration of a pulse train (also a round) is about $256ns$. A phase modulator (PM, Photline 10GHz) is employed to encode a random phase shift, $0$ or $\pi$, on each pulse.

On receiver side, an interferometer with variable delays is constructed in order to perform different interference measurements, as shown in Fig.~\ref{Fig:Setup}. The required delay time is a discrete value in $\{2ns,4ns,\cdots,254ns\}$. Seven delay gates, denoted as $DG_i$ for $i\in\{1,\cdots,7\}$ with a fixed delay fibre that can delay an optical pulse for a time of $2^i\times 2ns$, are arranged to achieve the 128-value dynamic delays. The length of delay path is carefully adjusted to ensure that pulses with and without delay can overlap well. Meanwhile, the coupling efficiency of every pair of collimators is above $90\%$ to ensure the intensity of pulses undergoing different numbers of delay gates as closely as possible. The seven delay gates are controlled by a 7-bit random number. By making use of these delay gates, we realize any dynamic discrete delay time from the set of $\{0 ns, 2 ns, 4 ns\cdots, 254 ns\}$, which includes all the required values of our experiment.

Each delay gate is constructed by a pockels cell, a fibre with fixed length and two polarizing beam splitters(PBS) as input port and output port, as shown in Fig.~\ref{Fig:Setup}. The pockels cell contains two RTP crystals. Each pockels cell is controlled by a customized high voltage pulse generator to achieve a fast switch between 0 V and half-wave voltage which is around 2100 V. For each one of delay gates, if the control bit is 0, the pockels cell will not affect the arrived pulses which will lead to pass through the output PBS and no delay happened.
If the control bit is 1, the pockels cell will be driven by the half-wave voltage to convert the polarization state of arrived photons from horizontal ($|H\rangle$) to vertical ($|V\rangle$). Thus the pulses will be reflected by the output PBS and propagate through the delay fibre. The delay gate of 2 ns ($DG_1$) is obtained by a free-space optical link of $\sim0.6$ m. The other delay gates of longer than 2 ns are obtained by the fibres with appropriate lengths. After reflected by the input PBS, the pulses pass through the half-wave voltage pockels cell again leading to convert the polarization back to $|H\rangle$ and pass through the output PBS. These seven delay gates are distributed in both arms of the interferometer to balance the transmittance. Some PBSs are shared between two neighbour gates.

After interfered at a fibre beam splitter (FBS), the pulses are detected by two custom up-convert single photon detectors\cite{Shentu:UpConvertion:2013} (SPD). By interacted with $1950nm$ pump laser in a PPLN waveguide, 1550 nm single photons are up-converted to 863 nm and then detected by commercial Si-based single photon detectors. The output of the SPDs are recorded by a high-speed and high-accuracy time-to-digital converter (TDC). The overall detection efficiency of the up-convert detector is around $10\%$, and the dark count is lower than 200cps.

\begin{figure}[!t]\center
\resizebox{16cm}{!}{\includegraphics{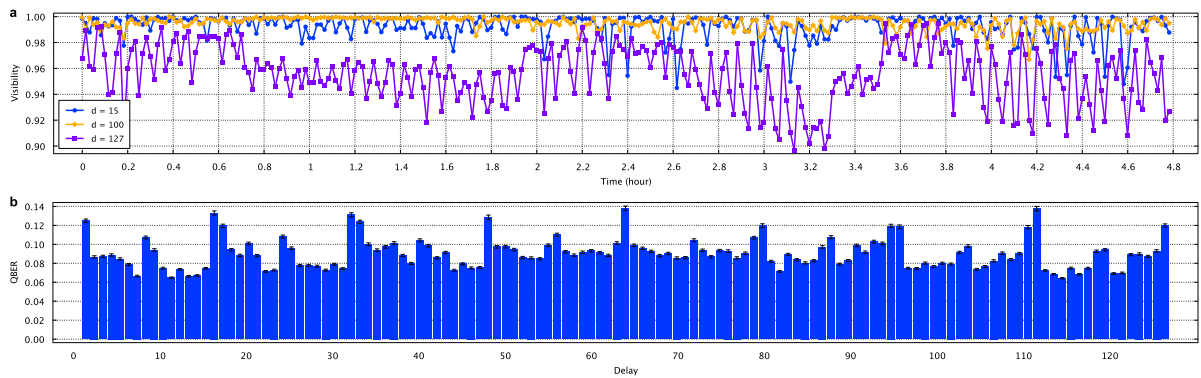}}
\caption{\textbf{a}, The phase stability with active phase locking over several hours. For most values of $d$, for example $d=15$ and $d=100$, the visibility is maintained above $96\%$. For certain delay such as $d=128$, the relative phase becomes a little more unstable. \textbf{b}, $e_{bit}^{(r)}$ over 128 logical delays. The mean error rate is $8.9\%$. In our experimental configuration of RRDPS QKD, the key rate keeps positive for each delay.}
\label{Fig:Result}
\end{figure}

In this experiment, we have realized a dynamic interferometer with 128 delays on measurement site. All of these possible interferometers require stability in the sub-wavelength order to perform high visibility interference measurements. To suppress mechanical vibration and temperature drift from the optical table and air, we employ a frame with thermal insulation and the high-damping material to envelop the interferometer. With these passive phase stabilisation methods, 128 kinds of unequal-arm interferometers can maintain a visibility of higher than $96\%$ for a time period in the order of ten seconds, depending on the delay $d$. The residual phase instability is mainly due to the drift of the central wavelength of the laser. Therefore, to implement a complete RRDPS QKD experimental demonstration, an active phase-locking technique is required. On sender side, an additional path without modulation of AM and PM, which is named phase-locking light with a relatively higher intensity of about 60 million photons per second, is introduced by an FBS and an optical switch. The phase-locking light is switched on for 340 ms per second is used for phase-locking to calibrate the interferometers. The rest 660 ms per second is used for QKD. During phase-locking, the 128 delays are traversed by turning on the specific Pockels Cells. $PM_2$ is deployed to adjust the relative phase between two arms of the interferometers. For each delay, the optimal compensate voltage of $PM_2$ is measured and recorded by a custom Field Programmable Gate Array (FPGA) in the control board (see Methods). The recorded compensate voltages are used to maintain the relative phase unchanged against different delay selections and central wavelength drift. With the active phase-locking technique, the visibility of most interferometers can be maintained over $96\%$ for hours simultaneously, as shown in Fig. \ref{Fig:Result}a.

\begin{table}
\centering
\arraycolsep=9pt
\renewcommand{\arraystretch}{1.5}
\caption{The experimental experimental parameters and results of our RRDPS experiment.}
\label{Tab:Result}
\begin{tabular}{ccccccccc}
  \hline
 Total rounds& Sifted key & Q&$\nu$ & $e_{bit}^{(mean)}$ & $L$ & Final key length& \\
  \hline
  103,679,400&675,937 & 0.00652&0.8 & 8.9\% & 128 & $2.441\times10^{5}$\\
  \hline
\end{tabular}
\end{table}

In this letter, the experiment is demonstrated through a 1 km fibre link. The intensity of each pulse is $0.00625$ and the total intensity of the $L$-pulse train is $0.8$. The repetition rate of pulse trains is 10 kHz. With 34\% of time used for calibration, 6600 pulse trains are transmitted per second. In the experiment, we obtain $675,937$ bits sifted keys in total. The bit flip error rates with different random delays $r$, $e_{bit}^{(r)}$, are shown in Fig.\ref{Fig:Result}$\mathbf{b}$. Due to the unbalance of the two arms of interferometer, there are fluctuation of the bit error rates with different delays. Interestingly, the bit error rate depends on the delay $r$ and the overall error rate is $8.9\%$. The total number of the rounds sent is $103679400$ and finally $2.441\times10^{5}$ bits of security key is generated in $15709$ seconds. Therefore, the key generation rate is given by $15.54$ bps.

Recently, another RRDPS QKD protocol is proposed and experimentally demosntrated\cite{guan2015PassiveRRDPS}. Difference from the original RRDPS protocol,
the random delay is chosen passively. A full scale comparison between two protocols is an interesting prospective project.


\begin{methods}

\textbf{Privacy amplification estimation.}
Our postprocessing essentially follows the improved result\cite{zhang2015round} as well as the original one\cite{sasaki2014practical}. The phase error rate of the RRDPS QKD¡¡protocol is solely determined by the source preparation. To be more specific, the leaked information Eve can obtained is upper-bound by the photon-number of the state Alice prepares in an $L$-pulse train\cite{sasaki2014practical}. In the case that the photon-number in an $L$-pulse train is a constant, a tighter bound of the phase error rate can be obtained by considering more details of the source\cite{zhang2015round}.
\begin{equation}\label{eq:Neph}
\begin{aligned}
e_{\textrm{ph}}^n &= \frac{1 - \left(1 - 2/L\right)^{n}}{2}, \\
\end{aligned}
\end{equation}
where $n$ is the total number of photons in the $L$-pulse train.

In the experiment, the weak coherent source is used as the source whose photon-number follows a Poisson distribution. When considering the worst case, we assume that Bob only receives high-photon number states, whose photon number is bigger than $n_{th}$, and losses always come from all the low-photon number states, whose photon number is smaller than $n_{th}$. Then $QH_{ph}$ , the cost of privacy amplification per round, is given by
\begin{equation}\label{nodecoy}
\begin{aligned}
QH_{PA}&\le Q_{n_{\textrm{th}}}H(e_{\textrm{ph}}^{n_{th}}) + \sum_{n = n_{th} + 1}^\infty e^{-\mu}\frac{(\mu)^n}{n!}H(e_{\mathrm{ph}}^n),\\
Q_{n_{\textrm{th}}}&=Q-\sum_{n = n_{th} + 1}^\infty e^{-\mu}\frac{(\mu)^n}{n!}.
\end{aligned}
\end{equation}
where $e_{\textrm{ph}}^{n_{\textrm{th}}}$ and $e_{\mathrm{ph}}^n$ can be upper-bounded according to Eq.~\eqref{eq:Neph}. And $n_{\textrm{th}}$ is the threshold photon number satisfy the total gain, $Q$, can be determined by contributions of the terms whose photon numbers are at least $ n_{\textrm{th}}$.

\textbf{Active phase stabilization.} The output intensity of one port of a Mach-Zehnder interferometer can be expressed as $\frac{I}{2}(1+\cos{\varphi_r})$, where $I$ is the input power and $\varphi_r$ is the relative phase between two paths for each delay $r$. The active phase stabilization technique is aimed to measure the value of $\varphi_r$ at phase-locking stage and compensate it by using a PM at QKD stage. To realize the measurement, the PM is set to apply extra phases of $\varphi_{ext}^{k}=\frac{2k}{N}\pi$ where $k\in{0, 1, ..., N}$ sequentially, corresponding to output intensity of $I_{k}=\frac{I}{2}(1+\cos(\varphi_r + \varphi_{ext}^{k}))$. For each voltage, the counting rate of two single photon detectors, $C_1^k$ and $C_2^k$ are recorded by FPGA. To determine $\varphi_r$, a least squares method is used to achieve the minimal of $S(\varphi_r)=\sum_k{(I_k-\frac{C_1^k}{C_1^k+C_2^k})^2}$. $N$ is chosen to be 4 in this experiment. Due to the imperfection, such as nonlinear of the voltage-phase mapping and fluctuations of photon number counting, the calculated compensate voltage of PM is usually not precise enough. The compensate voltage is then searched around the calculated value for a few steps to obtain a more suitable one. Finally, a lookup table is built to store the compensate voltages for each delay.

\textbf{Time sequence control.} The setup of Alice's side and Bob's side are controlled by two independent FPGA based on integrated control boards, $CB_a$ and $CB_b$. The phase-locking light is switched on for 340ms as the phase-locking stage. During this stage, Bob traversals each value of $r$ by switch on specific delay gates to measure the compensation phase $\varphi_r$. After phase-locking, Alice switches photon source to pulse mode which is modulated by AM and PM. The AM is directly controlled by amplified RF signal output from Arbitrary Waveform Generators(AWG), which is synchronized by $CB_a$, to generate pulses separated by 2ns with the FWHM of 300ps. Random numbers are generated by a QRNG module on $CB_a$ and transmitted to the AWG to compile as PM controlling waveform. The repetition rate of round is 10kHz in our experiment, thus phase encoded 128-pulse trains are produced per $100\mu s$. The start of rounds for AWG and $CB_b$ are triggered by $CB_a$ separately. At the beginning of per round, 7 bits of quantum random number generated by QRNG module on $CB_b$ are consumed to determine the delay $r$. Specific delay gates are turned on to construct the time delay. And $PM_2$ is set to the related voltage stored in the look-up table. Due to the slow rising time of the pockels cells, it takes $90\mu s$ to wait before trigger Alice's pulses to ensure the delay gates are precisely switched before. Finally, the trigger signal and the output of up-conversion detectors are recorded by a TDC on $CB_b$.

\end{methods}

\textbf{\subsection*{Acknowledgments}}
We acknowledge C.~Liu and X.~Han for their insightful discussions. This work has been supported by CAS Center for Excellence and Synergetic Innovation Center in Quantum Information and Quantum Physics, Shanghai Branch, University of Science and Technology of China, by the National Fundamental Research Program (under grant no. 2011CB921300 and 2013CB336800), and by the National Basic Research Program of China Grants No.~2011CBA00300 and No.~2011CBA00301, and the 1000 Youth Fellowship program in China.
\textbf{\subsection*{Author Contributions}}
All authors contributed extensively to the work presented in this paper.
\textbf{\subsection*{Author Information}}
The authors declare no competing financial interests.

\bibliographystyle{apsrev4-1}
\bibliography{bibRRDPS}
\clearpage

\end{document}